\documentclass[a4paper,11pt]{article}
\usepackage{pos}

\usepackage{algorithmic}
\usepackage{array}
\usepackage{graphicx}
\usepackage[absolute,overlay]{textpos}

\title{Implementation of the conjugate gradient algorithm for heterogeneous systems}

\author*[a]{Salvatore Cal\`i}
\author[a]{William Detmold}
\author[b]{Grzegorz Korcyl}
\author[c]{Piotr Korcyl}
\author[a]{Phiala~Shanahan}

\affiliation[a]{Center for Theoretical Physics, Massachusetts Institute of Technology, Cambridge, MA 02139, USA}

\affiliation[b]{Institute of Applied Computer Science, Jagiellonian University, ul. prof. {\L}ojasiewicza 11, 30-348 Krak{\'o}w, Poland}

\affiliation[c]{Institute of Theoretical Physics, Jagiellonian University, ul.
prof. {\L}ojasiewicza 11, 30-348 Krak{\'o}w, Poland}

\emailAdd{calis@mit.edu}
\emailAdd{wdetmold@mit.edu}
\emailAdd{grzegorz.korcyl@uj.edu.pl}
\emailAdd{piotr.korcyl@uj.edu.pl}
\emailAdd{pshana@mit.edu}

\abstract{Lattice QCD calculations require significant computational effort, with the dominant fraction of resources typically spent in the numerical inversion of the Dirac operator. One of the simplest methods to solve such large and sparse linear systems is the conjugate gradient (CG) approach. In this work we present an implementation of CG that can be executed on different devices, including CPUs, GPUs, and FPGAs. This is achieved by using the SYCL/DPC++ framework, which allows the execution of the same source code on heterogeneous systems.
\flushright{MIT-CTP/5348} 
}

\FullConference{%
 The 38th International Symposium on Lattice Field Theory, LATTICE2021
  26th-30th July, 2021
  Zoom/Gather@Massachusetts Institute of Technology
}

\begin{document}
\maketitle

\section{Introduction}
With the diversification of computing hardware, programming languages that allow software to be executed on any combination of devices, including Central Processing Units (CPUs), Graphics Processing Units (GPUs) and Field Programmable Gate Arrays (FPGAs), are very appealing. Such a framework gives freedom to the user to delegate the most demanding parts of a calculation to a specific accelerator, depending on the problem at hand. This mitigates the costs of different programming and optimization for each resource which is present on a supercomputer node \footnote{In this direction, it is worth mentioning that there exist programming models like \textit{Kokkos} and \textit{Raja}, which are C++ abstraction layers for performance portable parallel execution. Using specific features of these models, an algorithm can be mapped onto existing parallel programming languages and frameworks, like CUDA, OpenMP \cite{openmp}, HIP \cite{hip}, and SYCL/DPC++ \cite{sycl, reinders2020data}.}. This is the motivation for the standard SYCL/DPC++ (Data Parallel C++) (see~\cite{reinders2020data} for an introduction), and in this work we explore the possibility of using this framework for lattice QCD (Quantum Chromodynamics) applications. 

One of the most common problems in computational science and linear algebra is to solve systems of linear equations. For instance, in a typical lattice QCD calculation, most of the computing resources are spent in finding the solution $\psi$ of the following linear system:
\begin{equation}
D^{AB}_{\alpha\beta}(n,m)\psi^{B}_{\beta}(m) = \eta^{A}_{\alpha}(n),\quad A,B\in\lbrace 0,1,2\rbrace,\quad \alpha,\beta\in \lbrace 0,1,2,3\rbrace,
\label{eq:dirac_operator}
\end{equation} 
where $D$ is the Wilson-Dirac operator \cite{Wilson:1974sk} and $\eta$ is an arbitrary source fermion field. In Eq.~\eqref{eq:dirac_operator} $A,B$ are color indices, $\alpha,\beta$ are spin indices and $m,n$ represent two space-time coordinates within the lattice volume. To give a concrete example of the typical sizes, a lattice with moderately size volume $V=48^3\times 96$ would require the linear system~\eqref{eq:dirac_operator} to be solved for a matrix of size $(V\times 12)^2\approx 10^8\times 10^8$. As a consequence, solving \eqref{eq:dirac_operator} using a direct approach is not feasible and iterative methods for sparse matrices are employed to drastically reduce the computational effort and the storage needs. Examples of iterative methods used in lattice QCD computations are: conjugate gradient (CG), biconjugate gradient (BiCG), and the biconjugate gradient stabilized method (BiCGSTAB). The solver performance is often improved by preconditioning, such as Algebraic Multigrid (AMG), Incomplete Cholesky factorization (IC), Jacobi method, etc. Recently deep learning techniques have been investigated to design preconditioning matrices for such systems \cite{Xiao:MLP, Sappl:2019}. For a general introduction to iterative methods and preconditioning techniques we refer to Ref.~\cite{saad03:IMS}.

In Ref.~\cite{Korcyl:2020veo}, an OpenCL implementation of the CG algorithm for Xilinx FPGAs has been presented, showing that FPGAs can provide competitive performance for the inversion of the Dirac operator (around 607 GFLOPs running in single precision on a Xilinx U280 Alveo card). For recent progress in the FPGA optimized HPCG benchmark, we also refer to Ref.~\cite{zeni2021optimized}, where the authors consider solving a simple elliptic partial differential equation discretized with a 27-point stencil on a regular 3D grid using the CG algorithm.

The idea of this project is to explore the possibility that a single-source code can be used on different architectures for lattice QCD applications. Therefore, in this work we consider a single-node DPC++ implementation of the Conjugate Gradient algorithm applied to the Wilson-Dirac operator. This implementation is executed on different devices (CPUs, GPUs and FPGAs) and we test the performances in each case.

\section{Numerical details}
For these calculations, we explore lattice volumes ranging from $4^4$ up to $14^4$ and focus on the inversion of the standard Wilson-Dirac operator. The latter is known to satisfy the $\gamma_5$-hermiticity, $\gamma_5 D \gamma_5 = D^{\dagger}$, so to use the CG algorithm we first solve for $DD^{\dagger}$ (which is hermitian by construction) and then we multiply the solution by $D^{\dagger}$. In this context, the standard CG algorithm reads
\begin{algorithmic}
\STATE $\psi \gets \psi_0$
\STATE $r \gets \eta - D \psi$
\STATE $p \gets r$
\WHILE {$|r| \geq r_{\text{min}}$} 
	\STATE $r_{\text{old}} \gets |r|$
        \STATE $\alpha \gets \frac{r_{\text{old}}}{|D^{\dagger} p|}$
        \STATE $\psi \gets \psi + \alpha p$
	\STATE $r \gets r - \alpha D D^{\dagger} p$

	\STATE $\beta \gets \frac{|r|}{r_{\text{old}}}$
	\STATE $p \gets r + \beta p$
\ENDWHILE
\label{alg. cg},
\end{algorithmic}
where $\psi_0$ represents the initial guess. To access the elements of the Wilson-Dirac operator, we store the sparse matrix $D$ in the so-called coordinate format, i.e. $D$ is stored using three arrays: one array contains the value of the non-zero elements and the other two arrays the corresponding row and column indices. Such an approach is not optimal and does not take advantage of the diagonal structure of D, but it is easy to implement and it is used as a starting point of this exploratory study. Using the coordinate format, a pseudo-code for the sparse-matrix vector multiplication needed in CG reads

\begin{algorithmic}
 \FOR{$k = 1$, $k{+}{+}$, while $N_{\text{NZ}}$} \STATE {out[Row[k]] = out[Row[k]] + Val[k]*in[Col[k]];} \ENDFOR
 \label{alg. spmv},
\end{algorithmic}
\vspace{0.125cm}
where $N_{\text{NZ}}$ is the number of non-zero elements and \textit{in} and \textit{out} are the input and output vectors. Row, Col and Val are the three arrays used to store the sparse matrix (respectively the row and column indices and the corresponding value).

SYCL/DPC++ offers several ways to invoke a kernel and in this work we mainly use the so-called \textit{ND-range kernels}. This approach allows a kernel function to be invoked on each iteration of the task; moreover, the programmer has full control of the parallelism and has the possibility to split the global size of the problem into a desired number of smaller blocks, called \textit{workgroups}. Each workgroup can be interpreted as a 1, 2, or 3 dimensional block of threads and contains a set of work items that are mapped, e.g. to a core in a GPU. The workgroup size determines the occupancy of the compute units and to achieve optimal performances a tuning of the workgroup size is needed.
Using ND-range kernels, we have implemented in SYCL/DPC++ all the main operations that appear in the CG algorithm: the sparse-matrix vector multiplication (which represents the most demanding part in terms of computational cost), the dot product, and vector additions/differences. 

The code has been tested on three different types of hardware:
\begin{itemize}
\item CPUs: Intel(R) Xeon(R) Gold 5218 CPU @ 2.30 GHz
\item GPUs: Nvidia GeForce RTX 2080 Ti, Nvidia A100-PCIE-40GB
\item FPGAs: Intel Arria 10, Intel Stratix 10
\end{itemize}
and in the next section, we discuss the performance that we have obtained on these devices.

\section{Results}
We split this section into two parts: the first details the results obtained on CPUs and GPUs and the latter contains some preliminary results and considerations related to SYCL/DPC++ codes on FPGA hardware.

\subsection{CPUs and GPUs}
\begin{figure}[h]
\centering
\includegraphics[width=0.75\textwidth]{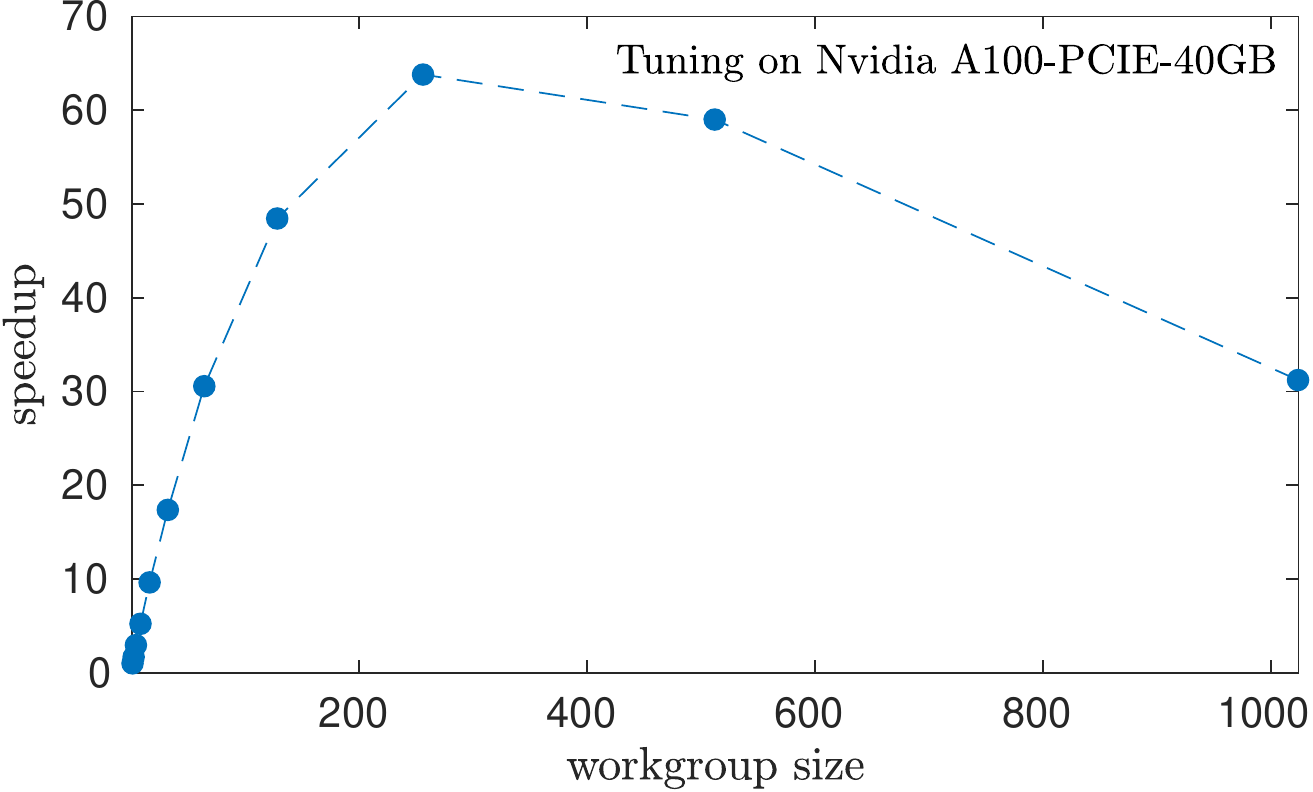}
\caption{Tuning of the optimal workgroup size on the Nvidia card A100-PCIE-40GB, performed for our implementation of the CG algorithm in SYCL/DPC++.}
\label{fig:tuning}
\end{figure}

\begin{figure}[h]
\centering
\includegraphics[width=0.75\textwidth]{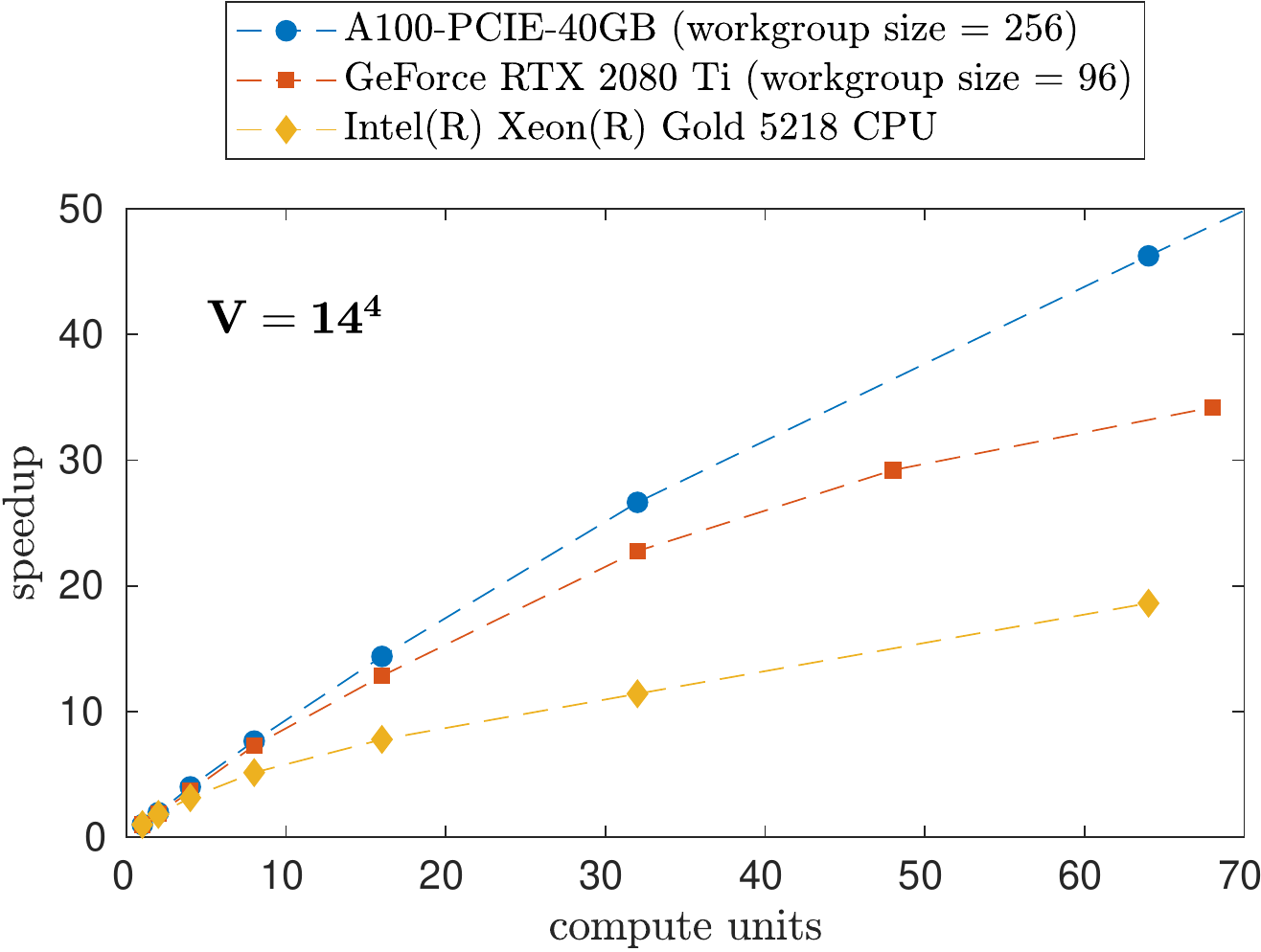}
\caption{Speedup study of our implementation of the CG algorithm in SYCL/DPC++, for a volume $V=14^4$ and with the Dirac operator stored in coordinate format. The reference point is the execution time obtained using a single compute unit.}
\label{fig:speedup}  
\end{figure}

Before showing the main results of this investigation, we highlight that on GPU hardware, in order to obtain the optimal performance, an initial tuning of the workgroup size is needed. For example, in Figure~\ref{fig:tuning} we report a study of the performance of the whole CG algorithm for a lattice volume $14^4$, keeping the number of compute units fixed and varying the workgroup size for the Nvidia A100-PCIE-40GB card.
In this case we find an optimal workgroup size of $256$. A similar study has been performed on Nvidia GeForce RTX 2080 Ti, for which we find it is more optimal to use a workgroup size of $96$. Once this tuning has been performed, we look at the speedup on CPUs and GPUs as a function of the compute units, as shown in  Figure~\ref{fig:speedup}. We observe in general a better scaling on NVIDIA GPUs and in particular for Nvidia A100-PCIE-40GB, than for the GeForce RTX 2080 Ti and the CPU architecture.

We also study the performances of the kernels in terms of GFlops/s, with special focus on the most demanding part, i.e., the sparse-matrix vector multiplication. As an approximation of the maximal theoretical performance that can be achieved on the hardware considered here, we compare the actual performances with the expectations of the naive roofline model~\cite{Williams09}. In this context, the maximal attainable performance $P$ (Flops/s) is given by
\begin{equation}
P = \min(\pi, \beta\times I),
\end{equation} 
where $\pi$ is the peak performance, $\beta$ (Bytes/s) is the peak bandwidth and $I$ (Flops/Bytes) is the operational intensity. Operations like sparse-matrix vector multiplications, dot products, and vector additions have low operational intensities and in these cases the kernel is said to be \textit{memory-bound} ($I<\pi/\beta$).

\begin{figure}[h]
\centering
\includegraphics[width=0.75\textwidth]{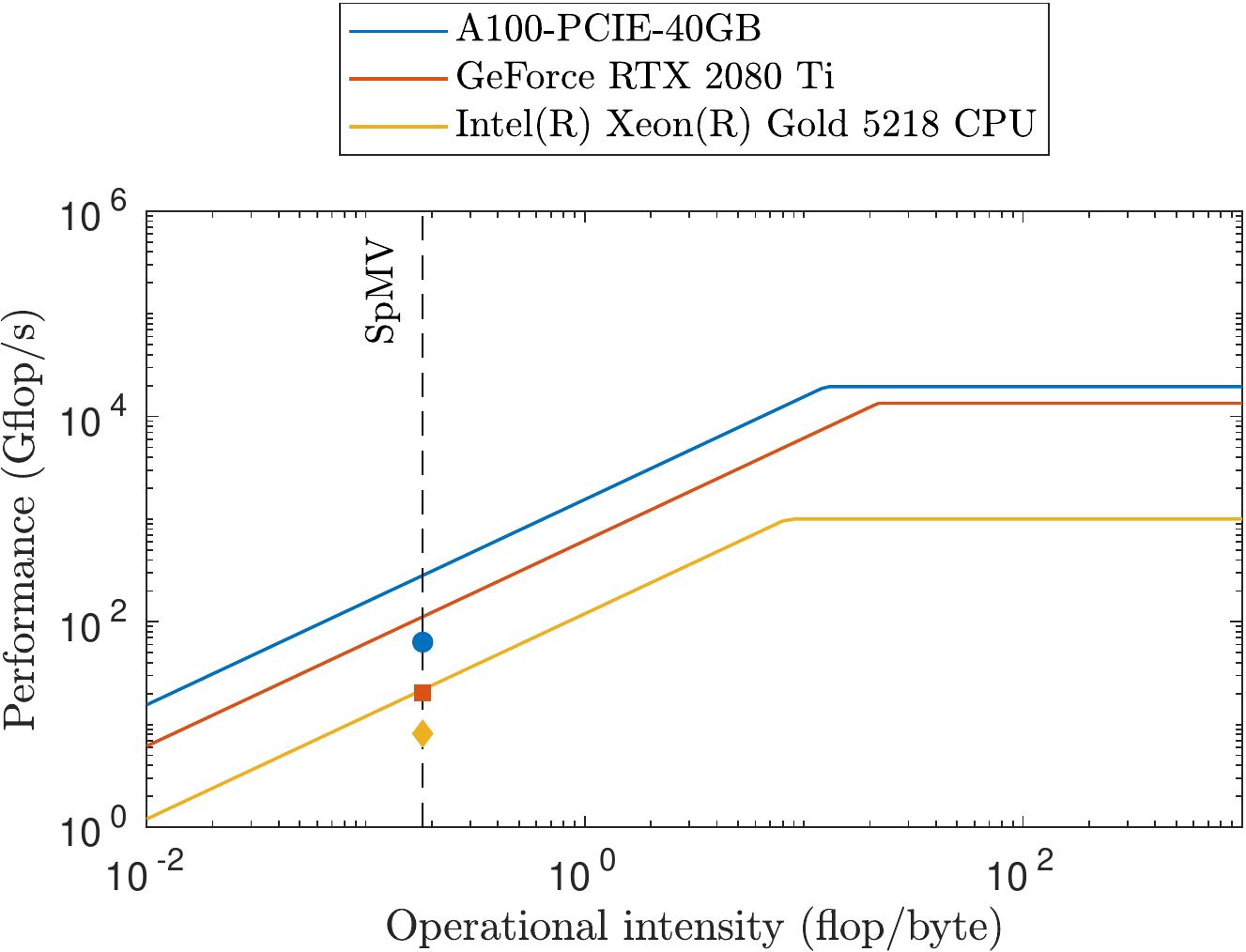}
\caption{Performance study of the sparse-matrix vector multiplication on different devices (points) and comparison with the naive roofline model (solid lines).}
\label{fig:roofline}
\end{figure}

In Figure~\ref{fig:roofline}, we show the performances that we obtain for the sparse-matrix vector multiplication, along with the theoretical limit of the roofline model.
As we can see from the figure, the maximum performance we can reach is around 65 GFlops/s on Nvidia A100-PCIE-40GB and, in general, for all the devices shown we observe that the actual performances are around $60\%-70\%$ of the theoretical ones.

\subsection{FPGAs}
The same SYCL/DPC++ code, tested on CPUs and GPUs as discussed above, has also been compiled and executed for Intel FPGAs. We find that the same source code is able to run on FPGA hardware, as we expect from the SYCL/DPC++ framework. However, we experience portability problems and using the code as it is leads to much worse performance on the FPGAs, typically below 1 GFlops/s. Therefore, we are currently far from realizing the idea of a single source code, running with good performance on all devices, and we are testing alternative kernels for execution on FPGAs. 

One idea is to consider single-task kernels, implementing loop unrolling mechanisms as described in the Intel oneAPI Github repository \url{https://github.com/oneapi-src/oneAPI-samples}. For kernels implementing simple functions, like vector addition, this approach (which consists of combining single-task kernels and the compiler directive ``\#pragma unroll'') seems very promising, giving rise to a significant performance improvement. However, we still face performance issues with the sparse-matrix vector multiplication kernel, probably because of the frequent memory accesses of the algorithm when the Dirac operator is stored in coordinate format. In the future, it could be interesting to test such kernels for a more suitable representation of the Dirac operator.

\section{Conclusions and outlook}
In these proceedings we have explored for the first time a SYCL/DPC++ implementation of the CG algorithm for the Wilson-Dirac operator. This framework allows a single-source application to be executed on different architectures, and we have tested our software on CPUs, GPUs, and FPGAs. Using the so-called ND-range parallel kernels and the Dirac matrix stored in coordinate format, we have seen that it is possible to obtain acceptable performances on CPUs and GPUs (around 65 GFlops/s on Nvidia A100-PCIE-40GB), that can be further improved designing algorithms more suitable for the Dirac operator. The same code also runs on FPGA hardware, but we observe worse performances, generally below 1 GFlops/s. Therefore, although the idea of having a single-source code to solve the Dirac equation running on different architectures is very appealing, at the moment we are far from this goal. More investigations are needed and in the near future we plan to consider different kernels and matrix representations that can speed up the execution times on FPGAs.

\section{Acknowledgments}
This work was supported by the Foundation for Polish Science grant no. TEAM/2017-4/39, by the Polish Ministry for Science and Higher Education grant no. 7150/E-338/M/2018, and by the Priority Research Area Digiworld under the program Excellence Initiative – Research University at the Jagiellonian University as well as by the U.S. Department of Energy, Office of Science, Office of Nuclear Physics, under grant contract numbers DE-SC0011090 and DE-SC0021006, and by the Carl G and Shirley Sontheimer Research Fund at MIT. WD is also supported by the SciDAC4 award DE-SC0018121. We gratefully acknowledge INTEL for providing access to FPGA hardware through the development sandbox \textit{Intel DevCloud}.

\bibliographystyle{JHEP}
\bibliography{biblio}


\end{document}